\documentclass[preprint,12pt]{elsarticle}
\usepackage{graphicx,amsmath}
\usepackage{xcolor}

\definecolor{mblue}{rgb}{.227,.324,.64}
\definecolor{mgreen}{rgb}{.063,.5,.25}
\definecolor{mred}{rgb}{.929,.133,.141}
\definecolor{mmagenta}{rgb}{.725,.325,.621}
\usepackage{amssymb}

\newcounter{bla}

\journal{Computer Physics Communications}

\begin{document}

\begin{frontmatter}



\title{An efficient finite-difference scheme for computation of electron states in free-standing and
core-shell quantum wires}


\author[a]{V. V. Arsoski\corref{author}}
\author[a,b]{N. A. \v{C}ukari\'c}
\author[a]{M. \v{Z}. Tadi\'c}
\author[b]{F. M. Peeters}

\cortext[author] {Corresponding author.\\\textit{E-mail address:} vladimir.arsoski@etf.bg.ac.rs}
\address[a]{School of Electrical Engineering, University of Belgrade, P.O. Box 35-54, 11120 Belgrade, Serbia}
\address[b]{Department of Physics, University of Antwerp, Groenenborgerlaan 171, B-2020 Antwerp, Belgium}

\begin{abstract}
The electron states in axially symmetric quantum wires are
computed by means of the effective-mass Schr\"odinger equation,
which is written in cylindrical coordinates $\varphi$, $\rho$, and
$z$. We show that a direct discretization of the Schr\"odinger
equation by central finite differences leads to a non-symmetric
Hamiltonian matrix. Because diagonalization of such matrices is
more complex it is advantageous to transform it in a symmetric
form. This can be done by the Liouville-like transformation
proposed by Rizea et al. (Comp. Phys. Comm. 179 (2008) 466-478),
which replaces the wave function $\psi(\rho)$ with the function
$F(\rho)=\psi(\rho)\sqrt{\rho}$ and transforms the Hamiltonian
accordingly. Even though a symmetric Hamiltonian matrix is
produced by this procedure, the computed wave functions are found
to be inaccurate near the origin, and the accuracy of the energy
levels is not very high. In order to improve on this, we devised a
finite-difference scheme which discretizes the Schr\"odinger
equation in the first step, and then applies the Liouville-like
transformation to the difference equation. Such a procedure gives
a symmetric Hamiltonian matrix, resulting in an accuracy
comparable to the one obtained with the finite element method. The
superior efficiency of the new finite-difference scheme ({\it
FDM}) is demonstrated for a few $\rho$-dependent one-dimensional
potentials which are usually employed to model the electron states
in free-standing and core-shell quantum wires. The new scheme is
compared with the other {\it FDM} schemes for solving the
effective-mass Schr\"odinger equation, and is found to deliver
energy levels with much smaller numerical error for all the
analyzed potentials. It also gives more accurate results than the
scheme of Rizea et al., except for the ground state of an infinite
rectangular potential in freestanding quantum wires. Moreover, the
${\cal PT}$ symmetry is invoked to explain similarities and
differences between the considered {\it FDM} schemes.
\end{abstract}

\begin{keyword}
quantum wire \sep finite difference \sep finite element \sep
discretization \sep eigenvalue \sep effective-mass

\end{keyword}

\end{frontmatter}


\section{Introduction}
\label{1}

Recent advances in nanowire (quantum wire) fabrication technology
have led to an increased interest in the {\it vapor-liquid-solid}
(VLS) method \cite{Wagner1964}. It is a bottom-up process, which
has been used to produce freestanding quantum wires
\cite{Duan2000}, core-shell quantum wires
\cite{Lucot2011,Dillen2014}, nanowire superlattices
\cite{Gudiksen2002}, branched nanowires \cite{Cheng2012}, etc. They
have been made out of various semiconductors, including III-V
compounds \cite{Lucot2011}, silicon \cite{Schmidt2009}, germanium
\cite{Wu2001}, and their alloys. The huge progress in the field
has been driven by actual and potential applications of nanowires
in electronics and photonics. For example, transistors
\cite{Karni2012}, photovoltaic devices \cite{Kim2012},
light-emitting diodes \cite{Tchernycheva2014}, lithium batteries
\cite{Chan2008}, and chemical and biological sensors
\cite{Xie2012} have all been realized using nanowires.

In addition to advances in production tools, the models of
electronic structure of quantum wires has substantially progressed
during time, both in increasing complexity and higher precision
\cite{Peelaers2010}. For example, {\it ab initio} methods are
currently able to predict experimental results with sub-meV
accuracy \cite{Peelaers2010}, but are overcomplex to use for large
wires. For the latter, however, use of the effective methods, such
as the effective-mass and ${\bf k}\cdot{\bf p}$ theories, may be
suitable \cite{Tadic1994,Sidor2005,Pistol2008,Kishore2012}. We
note that modeling of electronic structure is essentially
important to understand transport and optical properties of
nanostructures and nanodevices. Moreover, the electronic structure
models of quantum wires provide a reliable and an inexpensive way
to design quantum wire systems with specific properties.

A convenient model for the electron states in quantum wires which
are wider than about 2 nm is the effective-mass theory. It has the
form of the Schr\"odinger equation written for the case of
position dependent effective mass, and is able to capture the
essential physics of the electron states. In practice it usually
assumes that the confinement potential arises from a band offset
between different semiconductors, yet the eigenproblem is usually
only numerically solvable. For example, the wave function can be
expanded in a basis of analytical functions \cite{Taseli1997}. But
such an approach is known to produce dense Hamiltonian matrices,
and could have low accuracy of the wave functions around numerical
boundaries \cite{Cukaric2012}. An attractive alternative is the
finite-difference method ({\it FDM}) \cite{Galeriu2004}, which
employs a discretization of the wave function and its derivatives
on a grid \cite{Morton1976,Thomas1995}. Finite difference
approximations are usually of low order \cite{Thomas1995},
therefore the {\it FDM} delivers sparse matrices which could be
diagonalized extremely fast. As a matter of fact, the {\it FDM}
has been adopted to numerically solve various equations in physics
\cite{Schwigert1999, Zhao2007, Ibrahim1998}. For example, the Poisson equation and the
Schr\"odinger equation are solved together in the Hartree
calculation of exciton states by using the same {\it FDM}
discretization \cite{Stier2001}. The robustness of the {\it FDM}
has been an essential criterion for its frequent use to model
systems where the electrons are confined in more than one
dimension, quantum wires and quantum dots \cite{Stier2001}, for
example.

When applying the {\it FDM} to solve the Schr\"odinger equation, a
grid should be constructed with special care about the regions
close to the interfaces. It is not a difficult task when quantum
wires have axial symmetry, which allows reducing the eigenproblem
to the computation of matrix elements that depend on only the
$\rho$ coordinate of the cylindrical system. However, the
effective-mass Schr\"odinger equation contains a term proportional
to the first derivative of the wave function with respect to the
radius. Because of this term the finite-difference approximation
makes the Hamiltonian matrix nonsymmetric.

\begin{figure}
     \begin{center}
       \includegraphics[width=14 cm]{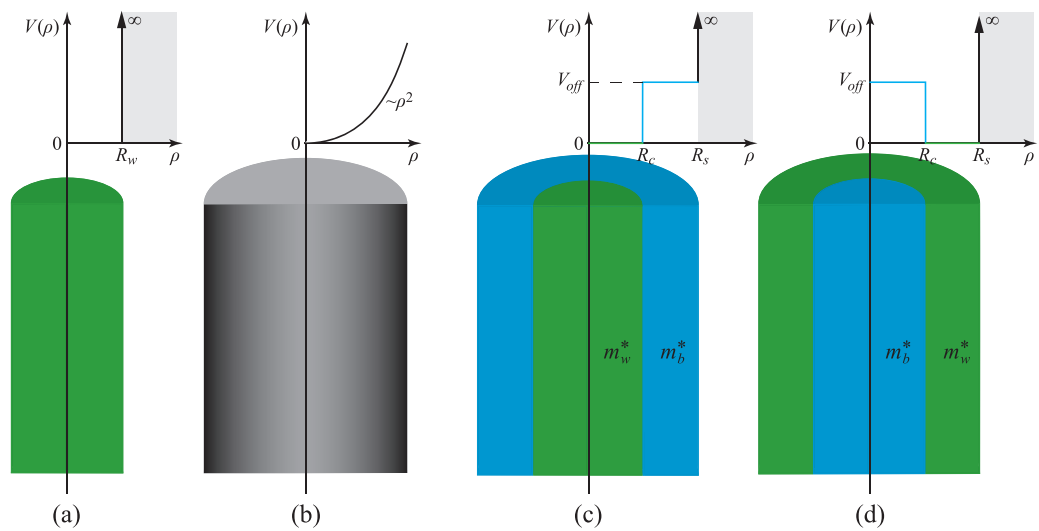}
    \caption{The considered potentials in the analyzed cylindrical quantum wires:
    (a) the infinite rectangular potential well in a free-standing
    quantum wire, (b) the potential of a linear harmonic oscillator,
    (c) the confining potential inside the core of a core-shell quantum wire,
    and (d) the confining potential inside the shell of a core-shell
    quantum wire.}
   \end{center}
\end{figure}

In this paper, we study how the {\it FDM} is used to solve the
effective-mass Sch\-r\"{o}\-di\-n\-ger equation for axially
symmetric potentials that appear in freestanding and core-shell
quantum wires. In the case of freestanding quantum wires, an
infinite rectangular potential and the potential of a 2D linear
harmonic oscillator are analyzed, shown schematically in
Figs.~1(a) and (b). Core-shell quantum wires are considered for:
(1) the type-Ic potential \cite{Pistol2008}, where the electron is
confined inside the core, and (2) the type-Is potential
\cite{Pistol2008}, which confines the electron in the shell. Both
analyzed potentials in core-shell quantum wires are assumed to
have stepwise variation with $\rho$, which is displayed in
Figs.~1(c) and (d). A few discretization {\it FDM} schemes are
constructed to solve the eigenproblem. {\it First}, the original
Schr\"odinger equation is discretized by central differences, and
it is demonstrated that the Hamiltonian matrix is asymmetric.
Furthermore, for computing the states of zero orbital momentum two
types of boundary conditions are tested and compared. {\it
Second}, the Schr\"odinger equation is transformed into another
equation \cite{Rizea2008} by the Liouiville-like ({\it LL})
transformation, which removes the problematic term from the
Hamiltonian. When the {\it LL}-transformed Schr\"odinger equation
is discretized by the {\it FDM}, the Hamiltonian matrix becomes
symmetric. However, the boundary condition at the inner boundary
is such that the wavefunctions are inaccurately computed close to
the origin. The {\it third} method is an approach developed by us,
which employs the finite-difference discretization of the original
Schr\"odinger equation, and subsequently applies the
Liouville-like transformation to the obtained difference equation.
This approach is novel to the best of our knowledge and is able to
solve the problem of insufficient accuracy of the solution of the
{\it LL}-transformed Schr\"odinger equation, and in the same time
delivers a symmetric Hamiltonian matrix. The accuracies of the
three discretization schemes are mutually compared for the
analyzed model potentials, and we compare the results with those
from the finite element method ({\it FEM}).

The paper is organized as follows. In Sec. II the discretization
schemes to solve the effective-mass Schr\"odinger equation for
quantum wires are presented. Sec. III contains the error analysis
on the example of a constant effective mass in the structure. Sec.
IV presents the results of our computations. We conclude in Sec V.

\section{The model of the electron states and the FDM schemes}
\label{2}

\subsection{The model}

We compute the electron states by using the effective-mass
Schr\"{o}dinger equation
\begin{equation}
    H_{3D}\Psi_{3D}({\bf r})=E\Psi_{3D}({\bf r}).
\end{equation}
Here, $H_{3D}$ denotes the single-band effective-mass Hamiltonian,
\begin{equation}
    H_{3D}=\frac{1}{2}{\bf p}\frac{1}{m^*({\bf r})}{\bf p}+V({\bf r}),
\end{equation}
where {$\it m^*(\bf r)$} is the position dependent electron
effective mass, {${\bf p}=-i\hbar\nabla$} is the canonical
momentum operator, and $V({\bf r})$ is the confining potential of
the electron. In an axially symmetric quantum wire grown along the
$z$ direction, $V({\bf r})=V(\rho)$ and $m^*({\bf r})=m^*(\rho)$,
where $\rho$ is the radius in the cylindrical coordinate system.
Thus the envelope function has the form
\begin{equation}
    \Psi_{3D}({\bf r})=Ce^{ik_zz}\Psi_{2D}(\rho,\varphi),
\label{psi3d:psi2d}
\end{equation}
where $C=const$, $\varphi$ is the polar angle, and $k_z$ is a good
quantum number representing the translational symmetry along the
$z$ direction. It reduces the complexity of the eigenproblem to
two coordinates, $\rho$ and $\varphi$,
\begin{equation}
    H_{2D}\Psi_{2D}(\rho,\varphi)=E\Psi_{2D}(\rho,\varphi),
\end{equation}
where
\begin{equation}
    H_{2D}=-\frac{\hbar^2}{2}
    \left[\frac{1}{\rho}\frac{\partial}{\partial\rho}\left(\frac{\rho}{m^*}\frac{\partial}{\partial\rho}\right)
    +\frac{1}{m^*}\frac{1}{\rho^2}\frac{\partial^2}{\partial\varphi^2}\right]
    +V(\rho).
\label{h2d}
\end{equation}
We can furthermore reduce the complexity by invoking the axial
symmetry,
\begin{equation}
    \Psi_{2D}({\bf r})=\frac{1}{\sqrt{2\pi}}e^{il\varphi}\psi(\rho),
\label{eq:psic}
\end{equation}
where $l$ denotes the orbital quantum number which represents
quantization of the $z$ projection of the electron angular
momentum $L_z$. The Schr\"odinger equation for $\psi$ is then
\begin{equation}
    -\frac{\hbar^2}{2}\left[\frac{1}{\rho}\frac{d}{d\rho}\left(\frac{\rho}{m^*}\frac{d\psi(\rho)}{d\rho}\right)
    -\frac{l^2\psi(\rho)}{m^*\rho^2}\right]
    +\left(V(\rho)+\frac{\hbar^2k_z^2}{2m}\right)\psi(\rho)=E\psi(\rho),
    \label{eq:seqr}
\end{equation}
which is written compactly as
\begin{equation}
    H\psi=E\psi.
\end{equation}
We will hereafter consider only the case $k_z=0$, since the
computation of the $k_z\neq 0$ states does not bring any
qualitative difference to the obtained results. With other words,
we model only the states of the quantum wire subband bottoms.
Expanding the first term on the left hand side of
Eq.~(\ref{eq:seqr}), the Schr\"odinger equation for $k_z=0$
becomes
\begin{eqnarray}
    &-&\frac{\hbar^2}{2}\left\lbrack\frac{1}{m^*}\frac{d^2\psi(\rho)}{d\rho^2}
    +\frac{1}{m^*}\frac{1}{\rho}\frac{d\psi(\rho)}{d\rho}+\frac{d}{d\rho}\left(\frac{1}{m^*}\right)\frac{d\psi(\rho)}{d\rho}
    -\frac{l^2\psi(\rho)}{m^*\rho^2}\right\rbrack\nonumber\\
    &+&V(\rho)\psi(\rho)=E\psi(\rho).
    \label{eq:derho}
\end{eqnarray}

The appropriate numerical domain to solve Eq.~(\ref{eq:derho}) has
the form of a cylinder of radius $R_{box}$ and is apparently
assumed to be of infinite height. The wave function is taken to be
zero at the cylinder surface, which corresponds to artificially
erecting the infinite potential barrier there. The boundary
condition should also be adopted at the inner boundary {$\rho=0$}.
It is derived by integrating Eq.~(\ref{eq:seqr}) over an
infinitesimally small region around the origin. This procedure
treats the $l=0$ states separately from the $l\neq 0$ states, and
results into: (1) $d\psi/d\rho|_{\rho=0}=0$ for $l=0$ (the von
Neumann boundary condition), and (2) $\psi(0)=0$ for $l\neq 0$
(the Dirichlet boundary condition). When $\psi$ is expanded in a
series close to $\rho=0$,
\begin{eqnarray}
    \psi(\rho)&=&\psi(0)+\frac{1}{1!}\frac{d\psi}{d \rho}\Big|_{\rho=0}\rho
    +\frac{1}{2!}\frac{d^2\psi}{d \rho^2}\Big|_{\rho=0}\rho^2+\cdots\nonumber\\
    &=&g_0+g_1\rho+g_2\rho^2+\cdots,\label{eq:exppsi}
\end{eqnarray}
the continuity of the wave function indicates that all the
derivatives $d^n\psi/d\rho^n$, $n\geq 1$, should be finite at
$\rho=0$. For $l=0$, $d\psi/d\rho|_{\rho=0}=g_1=0$, i.e. the
linear term is vanishing.

A direct approach to solve Eq.~(\ref{eq:derho}) is to discretize
it by central differences. However, discretization of the term
$(1/\rho)(d\psi/d\rho)$ makes the Hamiltonian a nonsymmetric
matrix. In order to make it symmetric, Rizea et al.
\cite{Rizea2008} proposed to replace $\psi$ with
\begin{equation}
    F(\rho)=\sqrt{\rho}\psi(\rho),
\label{eq:ccs}
\end{equation}
and to transform the Hamiltonian,
\begin{equation}
    \tilde{H}=\frac{1}{\sqrt\rho}H\sqrt\rho.
\label{eq:trans}
\end{equation}
Eqs.~(\ref{eq:ccs}) and (\ref{eq:trans}) keep the eigenstates
orthonormal,
\begin{equation}
    \int\displaylimits_0^{R_{box}} \psi_i^*\psi_j\rho d\rho=
    \int\displaylimits_0^{R_{box}}
    F_i^*F_jd\rho=\delta_{ij}.
\label{ort:norm}
\end{equation}
One may notice that the scalar product of different $F$ functions
is computed as in a rectilinear system (see Sec.~4.5 for a much
thorough explanation of this property).

The transformed Hamiltonian $\tilde{H}$ is Hermitian, as evident
from
\begin{eqnarray}
    &-&\frac{\hbar^2}{2}\left[\frac{d}{d\rho}\left(\frac{1}{m^*}\frac{d F(\rho)}{d\rho}\right)
    -\left(\frac{l^2-1/4}{m^*\rho^2}+\frac{1}{2\rho}\frac{d(1/m^*)}{d\rho}\right)F(\rho)\right]\nonumber\\
    &+&V(\rho)F(\rho)=EF(\rho).\label{eq:seqf}
\end{eqnarray}
We note that Eqs.~(\ref{eq:ccs}) and (\ref{eq:trans}) represent
the Liouville transformation when they are applied to the Bessel
differential equation. For a general case of confining potential
and effective-mass variation Eq.~(\ref{eq:seqf}) does not have the
Sturm-Liouville form, yet the problematic term $(1/\rho)(d/d\rho)$
is removed from the modified Hamiltonian, thus Eqs.~(\ref{eq:ccs})
and (\ref{eq:trans}) could be referred to as the {\it
Liouville-like (LL) transformation}.

To solve Eq.~(\ref{eq:seqf}), the boundary conditions for $F$
should be determined from the boundary conditions for $\psi$. The
series expansion of $F$ determined from Eqs.~(\ref{eq:exppsi}) and
(\ref{eq:ccs}) is
\begin{equation}
    F(\rho)=g_0\rho^{\frac{1}{2}}+g_1\rho^{\frac{3}{2}}+g_2\rho^\frac{5}{2}+\cdots.,\label{eq:expF}
\end{equation}
wherefrom it follows that the $n$-th order derivative of $F$ with
respect to $\rho$ is
\begin{equation}
    \frac{dF^n(\rho)}{d\rho^n}=C_0(n)g_0\rho^{-n+\frac{1}{2}}+C_1(n)g_1\rho^{-n+\frac{3}{2}}+C_2(n)g_2(z)\rho^{-n+\frac{5}{2}}+\cdots.\label{eq:dexpF}
\end{equation}
Here, $C_i(n)=\prod_{q=0}^{n-1} (i+1/2-q)$. For $l\neq0$, $g_0=0$
and the first and the second derivative of $F$ are finite at
$\rho=0$. However, for $l=0$ and at $\rho=0$ the von Neumann
boundary condition for $\psi$ becomes the Dirichlet condition for
$F$, i.e. {$\lim\limits_{\rho\to 0}F\sim\lim\limits_{\rho\to
0}\sqrt{\rho} \to 0$} (see Eq.~(\ref{eq:expF})). Since $g_0\neq0$,
it follows from Eq.~(\ref{eq:dexpF}) that the derivatives of $F$
diverge at $\rho=0$, i.e. $\lim\limits_{\rho\to 0}d^n
F/d\rho^n\sim\lim\limits_{\rho\to 0}{\rho^{-n+1/2}} \to\infty$.
Thus it is difficult to compute the wave function $\psi$ close to
$\rho=0$ by solving Eq.~(\ref{eq:seqf}). Hence, the eigenenergies
of the $l=0$ states are determined with a large numerical error.
In order to resolve this problem, Rizea et al. \cite{Rizea2008}
used an asymmetric 3-point formula which fits the second
derivative to an analytical solution. However, that approach
generates nonsymmetric Hamiltonian matrices, and could therefore
be impractical. We instead devise a scheme which constructs
symmetric Hamiltonian matrices, and has the same accuracy as
solving the original Schr\"odinger equation, which is elaborated
in Sec.~4.5.

\subsection{The discretization schemes}
\label{2.1}

To solve both Eqs.~(\ref{eq:seqr}) and (\ref{eq:seqf}), a grid
with uniformly distributed points,
\begin{equation}
    \rho_{j}=j\Delta\rho,\;\;\; j=0,1,...,N_{\rho},
\end{equation}
is formed in the numerical domain $\rho\in[0,R_{box}]$ ($\Delta
\rho$ is the step size). The difference equations are derived from
Eqs.~(\ref{eq:seqr}) and (\ref{eq:seqf}) by adopting the
approximation of central differences for $j=1,...,N_{\rho}-1$.

\vspace{0.5 cm}\noindent (1) By applying the central differences
to Eq.(\ref{eq:seqr}), we get
\begin{eqnarray}
    {\it S-FDM}:&-&\frac{\hbar^2}{2(\Delta\rho)^2}\left[\frac{1+\frac{1}{2j}}{m^*_{j+\frac{1}{2}}}\psi_{j+1}-\Bigg(\frac{1+\frac{1}{2j}}{m^*_{j+\frac{1}{2}}}\right.\nonumber\\
                &+&\left.\frac{1-\frac{1}{2j}}{m^*_{j-\frac{1}{2}}}+\frac{l^2}{j^2m^*_{j}}\Bigg)\psi_{j}+\frac{1-\frac{1}{2j}}{m^*_{j-\frac{1}{2}}}\psi_{j-1}\right]\nonumber\\
                &+&V_{j}\psi_{j}=E\psi_{j},\label{eq:dps}
\end{eqnarray}
which we call the {\it S-FDM} scheme, where the letter $S$
symbolizes a direct solution of the Sch\-r\-\"o\-di\-n\-ger
equation. A careful inspection of this equation shows that the
term proportional to the first derivative of $\psi$ with respect
to $\rho$ makes the Hamiltonian matrix asymmetric.

\vspace{0.5 cm}\noindent (2) Similar to the {\it S-FDM} scheme,
the central differences are also employed to discretize
Eq.~(\ref{eq:seqf}), which gives:
 \begin{eqnarray}
    {\it LL-FDM}:&-&\frac{\hbar^2}{2(\Delta\rho)^2}\left[\frac{1}{m^*_{j+\frac{1}{2}}}F^L_{j+1}-\Bigg(\frac{1+\frac{1}{2j}}{m^*_{j+\frac{1}{2}}}\right.\nonumber\\
    &+&\left.\frac{1-\frac{1}{2j}}{m^*_{j-\frac{1}{2}}}+\frac{l^2-1/4}{j^2m^*_{j}}\Bigg)F^L_{j}+\frac{1}{m^*_{j-\frac{1}{2}}}F^L_{j-1}\right]\nonumber\\
    &+&V_{j}F^L_{j}=EF^L_{j}.\label{eq:df0}
\end{eqnarray}
We refer to Eq.~(\ref{eq:df0}) as the {\it LL-FDM} scheme, and it
is straightforward to show that the constructed Hamiltonian matrix
is symmetric. But, as will be demonstrated below, the computation
of the $l=0$ states suffers from a low accuracy.

\vspace{0.5 cm}\noindent (3) In order to resolve the problems of
the {\it S-FDM} and {\it LL-FDM} discretization schemes, we
developed the third scheme. It replaces $\psi_{j}$ in
Eq.~(\ref{eq:dps}) with $F^{DL}_{j}/\sqrt{j\Delta\rho}$, and then
multiplies the difference equation by $\sqrt{j\Delta\rho}$,
\begin{eqnarray}
    {\it DLL-FDM}:&-&\frac{\hbar^2}{2\Delta\rho)^2}\left[\frac{1}{m^*_{j+\frac{1}{2}}}\frac{j+\frac{1}{2}}{\sqrt{j(j+1)}}F^{DL}_{j+1}-\Bigg(\frac{1+\frac{1}{2j}}{m^*_{j+\frac{1}{2}}}\right.\nonumber\\
    &+&\left.\frac{1-\frac{1}{2j}}{m^*_{j-\frac{1}{2}}}+\frac{l^2}{j^2m^*_{j}}\Bigg)F^{DL}_{j}+\frac{1}{m^*_{j-\frac{1}{2}}}\frac{j-\frac{1}{2}}{\sqrt{j(j-1)}}F^{DL}_{j-1}\right]\nonumber\\
    &+&V_{j}F^{DL}_{j}=EF^{DL}_{j}.\label{eq:df}
\end{eqnarray}
Because the discretization is done by adopting the {\it LL}
transformation at the discrete points $\psi_{j}$, we name it the
{\it DLL-FDM} scheme. Such constructed Hamiltonian matrix is
symmetric, hence its diagonalization bring in all the real
eigenvalues.

\vspace{0.5cm}\noindent (4) {\it R-FDM}: Yet another scheme is
proposed by Rizea et al. to solve the problem of a low accuracy of
the {\it LL-FDM} scheme \cite{Rizea2008}. It adapted the solution
for an arbitrary potential to the solution in an infinite
rectangular potential well. We refer to this discretization scheme
as the {\it R-FDM}.

\subsection{The boundary conditions}
\label{2.2}

In order to solve the derived difference equations, conditions
which are adopted at the outer boundary are either
$\psi_{N_\rho}=0$ or $F_{N_\rho}=0$. They are implemented by
removing the column $j=N_{\rho}$ from the system. On the other
hand, the values of $\psi$ and $F$ at the inner boundary are
implemented by either adding an equation for $j=0$, or modifying
an equation for $j=1$, as will be explained separately for each
discretization scheme.

\vspace{0.5 cm} \noindent {\it S-FDM:} When {$l\neq 0$} there is a
singular term {$l^2/\rho^2$} in the Schr\"odinger equation, and
physically we have to take $\psi_0=0$ at $\rho=0$. For this case,
no equation should be added or modified in the system in
Eq.~(\ref{eq:dps}). On the other hand, for $l=0$ the boundary
condition at the inner boundary can be adopted in two forms.

\vspace{0.5cm} \noindent ($1$) {\it The extended boundary
condition.}  Let us consider how the radial part of the Laplacian
acts on {$\psi(\rho)$},
\begin{eqnarray}
    \Delta_{_{\rho}}\psi(\rho)=\frac{d^2\psi(\rho)}{d\rho^2}+\frac{1}{\rho}\frac{d\psi(\rho)}{d\rho}.\label{eq:Lrad}
\end{eqnarray}
The limiting value of Eq.~(\ref{eq:Lrad}) at {$\rho\to0$} could be
found by applying the l'Hospital rule to the second term in this
equation, which gives
\begin{equation}
    \lim\limits_{\rho\to0}\Delta_{_{\rho}}\psi(\rho)=\lim\limits_{\rho\to0}\frac{d^2\psi(\rho)}{d\rho^2}
    +\lim\limits_{\rho\to0}\frac{d\psi(\rho)/d\rho}{\rho}=2\lim\limits_{\rho\to0}\frac{d^2\psi(\rho)}{d\rho^2}.\label{eq:Lradl}
\end{equation}
For $l=0$, this equation is replaced in Eq.~(\ref{eq:derho}), and
is discretized by assuming $\psi_{-1}=\psi_{1}$ and
$m_{-1/2}=m_{1/2}$. It leads to an additional equation for $j=0$
\cite{Rizea2008},
\begin{equation}
    -\frac{\hbar^2}{2}\Bigg\{ \frac{2}{(\Delta\rho)^2}\Bigg[\frac{\psi_{1}}{m^*_{\frac{1}{2}}}-\left(\frac{1}{m^*_{\frac{1}{2}}}+\frac{1}{m^*_{-\frac{1}{2}}}\right)\psi_{0}
    +\overbrace{\frac{\psi_{-1}}{m^*_{-\frac{1}{2}}}}^{{\frac{\psi_{1}}{m^*_{\frac{1}{2}}}}}\Bigg]\Bigg\}
    +V_{0}\psi_{0}=E\psi_{0}.\label{eq:dr0}
\end{equation}
This boundary condition requires extending the $\rho$ axis to a
range where $\rho<0$, hence we call it the extended boundary
condition (hereafter abbreviated by $e$).

\vspace{0.5cm}\noindent ($2$) {\it The restricted boundary
condition.} The $e$ boundary condition has a drawback that the
$\rho$ axis is artificially extended to a range where $\rho$ is
not defined. The much simpler condition,
\begin{equation}
        \psi_{0}=\psi_{1},\label{eq:bcr}
\end{equation}
respects the requirement $\rho\geq 0$, and is therefore called the
restricted boundary condition (abbreviated by $r$). The
application of this boundary condition makes the equation for
$j=0$ redundant, which reduces the order of the Hamiltonian matrix
by unity.

For $l=0$, application of the restricted boundary condition to
Eq.~(\ref{eq:dps}) changes the equation for $j=1$,
\begin{equation}
    -\frac{\hbar^2}{4}\frac{3}{(\Delta\rho)^2m^*_{\frac{3}{2}}}\left(\psi_{2}-\psi_{1}\right)
    +V_{0}\psi_{0}=E\psi_{0},\label{eq:bcrp}
\end{equation}
whereas the equation for $j=0$ is superficial and is not added to
the system in Eq.~(\ref{eq:dps}). Hence, the matrices constructed
by means of the ${\it Sr}-{\it FDM}$ scheme for $l=0$ and $l\neq0$
are of equal order. On the other hand, the order of the
Hamiltonian matrix discretized by the ${\it Se}-{\it FDM}$ scheme
for $l=0$ is larger by unity with respect to the $l\neq 0$ case.

\vspace{0.5cm}\noindent {\it LL-FDM:} The boundary condition for
the function $F$ which should be supplied to the {\it LL-FDM}
difference equation is
\begin{equation}
     F^L_{0}=0,
\label{fl0}
\end{equation}
irrespective of the value of $l$.

\vspace{0.5cm}\noindent {\it DLL-FDM}: In the {\it DLL-FDM} scheme
$F^{DL}_{0}=0$ is substituted into Eq.~(\ref{eq:df}), but only
after the limiting value for $j=1$ is found,
\begin{eqnarray}
    \frac{-\hbar^2}{2(\Delta\rho)^2}\lim\limits_{j\to1}\left[\frac{1}{m^*_{j-\frac{1}{2}}}\frac{j-\frac{1}{2}}{\sqrt{j(j-1)}}F^{DL}_{j-1}\right]
    &=&\frac{-\hbar^2\sqrt{\Delta\rho}}{4m^*_{\frac{1}{2}}(\Delta\rho)^2}\lim\limits_{j\to1}\left[\frac{F^{DL}_{j-1}}{\sqrt{(j-1)\Delta\rho}}\right]\nonumber\\
    =\frac{-\hbar^2\sqrt{\Delta\rho}}{4m^*_{\frac{1}{2}}(\Delta\rho)^2}\psi_{0}
    &=&\frac{-\hbar^2\sqrt{\Delta\rho}}{4m^*_{\frac{1}{2}}(\Delta\rho)^2}\psi_{1}\nonumber\\
    =\frac{-\hbar^2\sqrt{\Delta\rho}}{4m^*_{\frac{1}{2}}(\Delta\rho)^2}\frac{F^{DL}_{1}}{\sqrt{\Delta\rho}}&=&\frac{-\hbar^2}{4m^*_{\frac{1}{2}}(\Delta\rho)^2}F^{DL}_{1}.\label{eq:df1}
\end{eqnarray}
Therefore, Eq.~(\ref{eq:df}) is for $j=1$ modified to,
\begin{equation}
    -\frac{\hbar^2}{2}\frac{1}{(\Delta\rho)^2}\left[\frac{3}{2\sqrt{2}}\frac{1}{m^*_{\frac{3}{2}}}F^{DL}_{2}-\frac{3/2}{m^*_{\frac{3}{2}}}F^{DL}_{1}\right]
    +V_{1}F^{DL}_{1}=EF^{DL}_{1}.\label{eq:dfm}
\end{equation}
This boundary condition obviously only affects the diagonal terms
in the matrix constructed from Eq.~(\ref{eq:df}), and therefore
the symmetry of the Hamiltonian matrix is preserved after the
application of the boundary condition. Note that the condition
$\psi_1=\psi_0$ is adopted in Eq.~(\ref{eq:bcr}), therefore the
boundary condition implemented in the {\it DLL-FDM} scheme is
equivalent to the one used in the {\it Sr-FDM} scheme to solve the
Schr\"odinger equation.

\section{The error analysis}
\label{3}

The error analysis of the different FDM schemes could be able to
explain differences between them. However, in order to simplify
this analysis, we assume that the effective mass is constant.

\vspace{0.5cm}\noindent {\it S-FDM}: The difference equation for
the wave function $\psi$ constructed from Eq.~({\ref{eq:dps}}) for
the case $m^*=const$ has the form:
\begin{equation}
    -\frac{\hbar^2}{2m^*}\left[\frac{\psi_{j+1}-2\psi_{j}+\psi_{j-1}}{(\Delta\rho)^2}+\frac{\psi_{j+1}-\psi_{j-1}}{2j(\Delta\rho)^2}-\frac{l^2}{j^2(\Delta\rho)^2}\right]+V_j\psi_j=E\psi_j.\label{eq:seq1p}
\end{equation}
The radial part of the Laplacian acting on $\psi(\rho)$ is
discretized such that
\begin{eqnarray}
    &&\frac{\psi_{j+1}-2\psi_{j}+\psi_{j-1}}{\Delta\rho^2}+\frac{\psi_{j+1}-\psi_{j-1}}{2j\Delta\rho^2}\nonumber\\
    &=&\left[\frac{1}{\rho}\frac{d}{d\rho}\left(\rho\frac{d\psi}{d\rho}\right)\right]_j\nonumber\\
    &+&\frac{(\Delta\rho)^2}{12}\left\{\frac{1}{\rho}\frac{d}{d\rho}\left[\rho\frac{d}{d\rho}\left(\frac{1}{\rho}\frac{d}{d\rho}\left(\rho\frac{d\psi}{d\rho}\right)\right)\right]\right\}_j+\cdots\nonumber\\
    &=&\left[\frac{1}{\rho}\frac{d}{d\rho}\left(\rho\frac{d\psi}{d\rho}\right)\right]_j+\vartheta(\Delta\rho^2).\label{eq:der2cyl}
\end{eqnarray}
Because the derivatives of $\psi(\rho)$ computed by means of the
{\it S-FDM} are finite, the {\it S-FDM}'s order is
$\vartheta(\Delta\rho^2)$.

\vspace{0.5 cm}\noindent {\it LL-FDM}: The {\it LL-FDM} equation
for $F(\rho)$ derived from Eq.~(\ref{eq:df0}) is
\begin{equation}
    -\frac{\hbar^2}{2m^*(\Delta\rho)^2}\left(F^L_{j+1}-2F^L_{j}+F^L_{j-1}-\frac{l^2-1/4}{j^2}F^L_j\right)+V_jF^L_j=EF^L_j.\label{eq:seq1f}
\end{equation}
The error of the {\it LL-FDM} discretization scheme can be
estimated by using \cite{Abram1972}
\begin{equation}
    \frac{F^L_{j+1}-2F^L_{j}+F^L_{j-1}}{\Delta\rho^2}=\left(\frac{d^2 F}{d\rho^2}\right)_j+\frac{(\Delta\rho)^2}{12}\left(\frac{d^4F}{d\rho^4}\right)_j+\frac{(\Delta\rho)^4}{360}\left(\frac{d^6F}{d\rho^6}\right)_j+\cdots.\label{eq:der2f}
\end{equation}
Because all the derivatives $d^n F/d\rho^n$ for $l\neq0$ are
finite in the whole domain, the second derivative is approximately
given as
\begin{equation}
    \left(\frac{d^2 F}{d\rho^2}\right)_j=\frac{F^L_{j+1}-2F^L_{j}+F^L_{j-1}}{\Delta\rho^2}+\vartheta(\Delta\rho^2),\label{eq:der2fe}
\end{equation}
where $\vartheta(\Delta\rho^2)$ is the error estimate. For $l=0$,
all the derivatives of the function {$F(\rho)$} are infinite at
{$\rho=0$}, which makes use of this discretization scheme
inconvenient in practice.

For $\rho=\Delta\rho$, from Eq.~(\ref{eq:dexpF}) it follows
\begin{equation}
    \frac{d^4 F}{d\rho^4}\Big|_{\rho=\Delta\rho}
    \approx C_0(4)\cdot g_0\Delta\rho^{-\frac{7}{2}}.
\end{equation}
Thus, the second derivative at the $j=1$ point is approximated as
\begin{equation}
    \left(\frac{d^2 F}{d\rho^2}\right)_{j=1}\approx
    \frac{F^L_{2}-2F^L_{1}+F^L_{0}}{\Delta\rho^2}-\frac{C_0(4)}{12}\frac{F^L_1}{\Delta\rho^2}+\vartheta(\Delta\rho^2).\label{eq:der2fe0}
\end{equation}
Hence, if the standard three-point discretization formula for the
second derivative is used at the grid points near the origin, $F$
is computed at $\Delta\rho$ with a much larger than
$\vartheta(\Delta\rho^2)$.

\vspace{0.5cm}\noindent {\it DLL-FDM}: Similar to the previous two
discretization formulas, the difference equation for $F$ according
to the {\it DLL-FDM} scheme is simply obtained from
Eq.~(\ref{eq:df}),
\begin{eqnarray}
    &-&\frac{\hbar^2}{2m^*(\Delta\rho)^2}\Bigg[\frac{j+\frac{1}{2}}{\sqrt{j(j+1)}}F^{DL}_{j+1}-\Bigg(2+\frac{l^2}{j^2}\Bigg)F^{DL}_{j}\nonumber\\
    &+&\frac{j-\frac{1}{2}}{\sqrt{j(j-1)}}F^{DL}_{j-1}\Bigg]+V_{j}F^{DL}_{j}=EF^{DL}_{j}.\label{eq:seq1m}
\end{eqnarray}

Because the transformation in Eq.~(\ref{eq:trans}) is applied to
the difference equation after the boundary conditions are
implemented, the error of the {\it DLL-FDM} discretization is of
the same order as the error of the discretization by the {\it
S-FDM} scheme. The eigenvalues obtained by means of the {\it
DLL-FDM} and {\it Sr-FDM} schemes are in fact equal, because the
{\it DLL-FDM} scheme is obtained by transformation of the {\it
Sr-FDM} equations (Eqs.~(\ref{eq:dps}) and (\ref{eq:bcr})).

The error of the {\it DLL-FDM} discretization is estimated from
the error of the finite difference approximation for the Laplacian
acting on $\psi=F/\sqrt{\rho}$ (see Eq.~(\ref{eq:seq1p})). When it
is multiplied by $\sqrt{j\Delta\rho}$, the error becomes of the
order $\sqrt{j\Delta\rho}\cdot\vartheta(\Delta\rho^2)$. It thus
ranges from $\vartheta(\Delta\rho^{2.5})$ at the grid points near
the origin to $\vartheta(\Delta\rho^2)$ at the grid points close
to the outer numerical boundary. Hence, the error of the {\it
DLL-FDM} approximation is smaller at the grid points which are
closer to the inner boundary. It thus indicates that the energy
levels computed by means of the {\it DLL-FDM} are more accurate
than those determined by the {\it LL-FDM} scheme, irrespective of
the value of $l$.

\section{The numerical results for quantum wire states}
\label{4}

We will use the mentioned discretization schemes to compute the
energy levels in quantum wires based on the GaAs and ${\rm
Al_{0.3}Ga_{0.7}As}$ materials, for the following potentials: (i)
the infinite rectangular potential well in a freestanding GaAs
quantum wire (see Fig.~1(a)), (ii) the potential of the linear
harmonic oscillator in a freestanding GaAs quantum wire, which is
illustrated in Fig.~1(b), (iii) the confinement inside the core of
the core-shell GaAs/${\rm Al_{0.3}Ga_{0.7}As}$ quantum wire (type
I-c quantum wire), which is shown in Fig.~1(c), and (iv) the
potential which confines the electrons inside the shell of the
core-shell ${\rm Al_{0.3}Ga_{0.7}As}$/GaAs quantum wire (type I-s
quantum wire), which is displayed in Fig.~1(d). The potential (a)
is usually employed to model states in freestanding quantum wires,
whereas the model (b) can be used to approximate potentials in
freestanding quantum wires which arise from the self-consistency
effects. The models (c) and (d) are employed for computations of
the states in core-shell quantum wires, and assume confinement in
the core and shell, respectively. The effective-mass Schr\"odinger
equation can be analytically solved for all these cases, thus the
exact energy levels are known, and the accuracy of the proposed
discretization schemes can be estimated for all four potentials.
The parameters of the materials are taken from
Ref.~\cite{Singh1993}.

\subsection{The infinite rectangular potential well in a free-standing quantum wire}
\label{4.1}

Some quantum wires are formed as freestanding, either by etching
\cite{Proetto1992, Youtsey1994} or the VLS technique \cite{Wang2001}. The
electron in them is in fact confined by a few eV large band offset
equal to the electron affinity. It is therefore localized almost
fully inside the wire, as in an infinitely deep axially symmetric
potential well. We assume that the potential inside the quantum
wire equals zero and that it is infinite outside the wire,
\begin{equation}
    V(\rho)=\begin{cases}
                0, & \rho\leq R_w \\
                \infty, & \rho> R_w
            \end{cases}.
\end{equation}
The Schr\"odinger equation for this case has the form of the
Bessel differential equation \cite{Abram1972}, whose solutions are
\begin{equation}
    E_{l,n}=\frac{\hbar^2(\alpha_{l,n})^2}{2m^*R_w^2},
\end{equation}
and
\begin{equation}
    \psi_{l,n}(\rho)=\frac{\sqrt{2}}{R_w}\frac{J_l\left(\alpha_{l,n}\rho/R_{w}\right)}{\left|J_{l+1}(\alpha_{l,n})\right|},\nonumber\\
    \label{eq:an1d}
\end{equation}
where {$\alpha_{l,n}$} is the {$n$}th zero of the Bessel function
of the first kind $J_l(x)$.

The computed wave functions for $l=0$ and $n$=1, 2, and 3 are
shown in Fig.~2. The dashed lines and the lines denoted by symbols
show the results of the {\it LL-FDM} and {\it DLL-FDM}
discretizations, respectively, which are compared with the
analytical solutions, displayed by solid lines. The wave functions
determined by the {\it DLL-FDM} scheme obviously nicely fits the
analytical solutions. On the other hand, the wavefunctions which
are computed by means of the {\it LL-FDM} exhibit a substantial
difference from the analytical solutions, especially at $\rho\ll
R_w$. As we previously inferred from Eq.~(\ref{eq:der2fe0}), the
approximation of the second derivative, given by
Eq.~(\ref{eq:der2fe}), is not valid at the grid points near the
origin. Therefore, the {\it LL-FDM} fails to properly reproduce
the wave function close to $\rho=0$. Nonetheless, as Fig.~2 shows
the wave functions of larger $n$ determined by the {\it LL-FDM}
differ less from the exact result. It is a consequence of the
increasing number of oscillations in higher-energy states, which
effectively narrows the region near the origin where
$\vert\psi\vert$ is large and where the {\it LL-FDM} discretized
function $F$ is computed with a large error. Notice that the value
of $\psi$ at $\rho=0$ could not be retrieved from our {\it LL-FDM}
calculation, since $F(\rho)/\sqrt{\rho}$ is indeterminate at
$\rho=0$.

\begin{figure}
     \begin{center}
       \includegraphics[width=8.6cm]{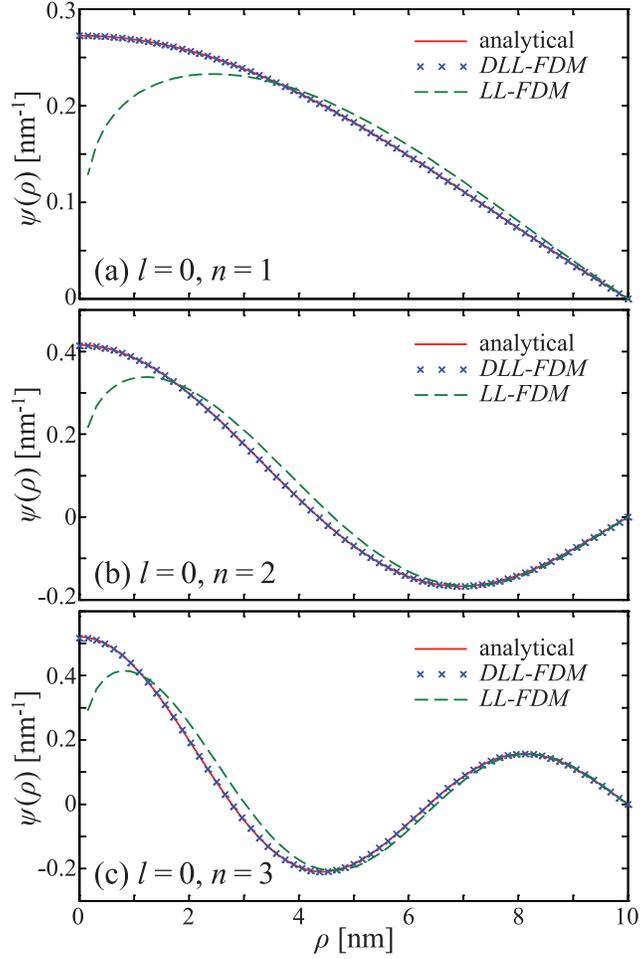}
    \caption{The radial parts of the wave functions
    calculated with the {\it DLL-FDM} scheme (blue symbols), the
    {\it LL-FDM} scheme (dashed green lines), and analytical wave
    functions (solid red lines) for {$l=0$}:
    (a) the ground {$(n=1)$} state,
    (b) the {$n=2$} state, and
    (c) the {$n=3$} state. Here $N_{grid}=64$.}
    \end{center}
\end{figure}

The {\it LL-FDM} and {\it DLL-FDM} schemes are also compared in
Fig.~3, where variation of the relative errors of the energy
levels with the number of grid points is shown. It is evident from
Figs.~3(a) and 3(b) that the relative error $\delta E=\Delta E /
E_{analytical}$ ($\Delta E=|E-E_{analytical}|$ is the absolute
error) of the energy levels computed by means of the {\it DLL-FDM}
is an order of magnitude smaller than the relative error of the
{\it LL-FDM} discretization. Also, the two discretization schemes
differ in the dependence of $\delta E$ on the level number $n$.
For a given $N_{grid}$, $\delta E$ determined by means of the {\it
DLL-FDM} increases with $n$. It is a consequence of the increasing
frequency of the wave function oscillations when $n$ increases,
which are difficult to accurately model with a small number of
grid points. Therefore, the relative error for higher states is
large when $N_{grid}$ is small. On the other hand, the relative
errors of the energy levels computed by means of the {\it LL-FDM}
decrease when $n$ increases, which is associated with the
demonstrated narrowing of the region where the deviation of $\psi$
from the accurate wave function is large (see Fig.~2).
Furthermore, as {\it a priori} expected, the relative errors of
the energy levels shown in both Figs.~3(a) and (b) decay when the
number of grid points increases. We include in Fig.~3(c) the
relative error of the energy levels determined by the {\it R-FDM}
scheme. The error of the {\it R-FDM} scheme is definitely lower
than of the other two for $n=1$, because the wavefunction is
conveniently fitted by a polynomial $P(\rho)$ of some degree
\cite{Rizea2008}. However, the use of the {\it R-FDM} scheme is
limited for higher energy states, because they are more
oscillatory. Then in order to improve the accuracy of the {\it
R-FDM} scheme, it is necessary to increase the order of the
polynomial which is used for the fitting procedure, which was
however not proposed in Ref.~\cite{Rizea2008}.

\begin{figure}
     \begin{center}
       \includegraphics[width=8.6cm]{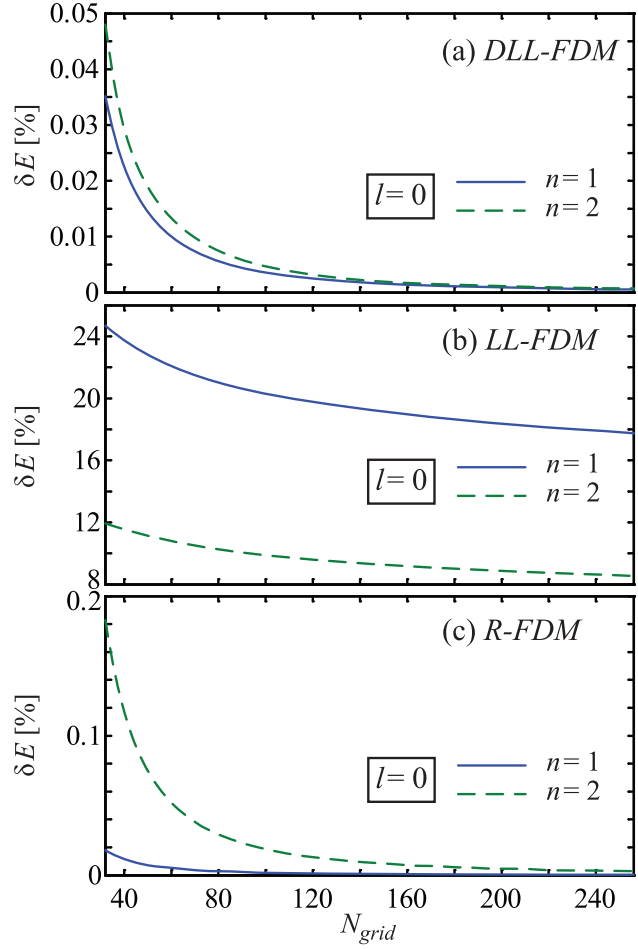}
    \caption{The relative errors of the $l=0$ energy
    levels, as function of the number of grid points
    for: (a) the {\it DLL-FDM}, (b) {\it LL-FDM}, and (c) {\it R-FDM}.
    $\delta E$ is shown for the ground $n=1$ state
    (solid blue lines) and the $n=2$ state (dashed green lines).}
    \end{center}
\end{figure}

For $l\neq 0$ the approximation of the second derivative by the
{\it LL-FDM} discretization given by Eq.~(\ref{eq:der2fe}) gives
results for the energy levels which are comparable to the {\it
DLL-FDM}. Nevertheless, the error of the {\it DLL-FDM}
approximation is smaller than the error of the {\it LL-FDM}, due
to a much better description of the wave function at the grid
points close to $\rho=0$. It is indeed demonstrated in Fig.~4,
where the absolute errors of the $n=1$, 2, and 3 energy levels are
shown as function of $N_{grid}$ for $l=1$. Here, not so large
difference between the results obtained by the two methods is
observed as in Fig.~3 for the $l=0$ state. This is mainly because
the Dirichlet boundary condition for $\psi$ at the inner boundary
is accurately reproduced by both the {\it LL-FDM} and the {\it
DLL-FDM} when $l=1$. Nonetheless, as the error analysis of the two
approximations showed, the {\it DLL-FDM} delivers the energy
states with a smaller absolute error than the {\it LL-FDM} scheme.
As apparent from Fig.~4, the errors of the $l\neq 0$ energy
levels determined by both discretization schemes increase with the
level number, opposite to what was previously shown in Fig.~3(b)
for the $l=0$ states found by the {\it LL-FDM}. Dotted lines in
Fig.~4 display the results of the {\it R-FDM} calculation for the
$l=1$ energy levels. Similar to the $l=0$ case shown in Fig.~3, a
smaller error is obtained for the $n=1$ state, but the error of
computation of the $n>2$ states by the {\it R-FDM} is larger than
of both the {\it DLL-FDM} and {\it LL-FDM} schemes.

\begin{figure}
     \begin{center}
       \includegraphics[width=8.6cm]{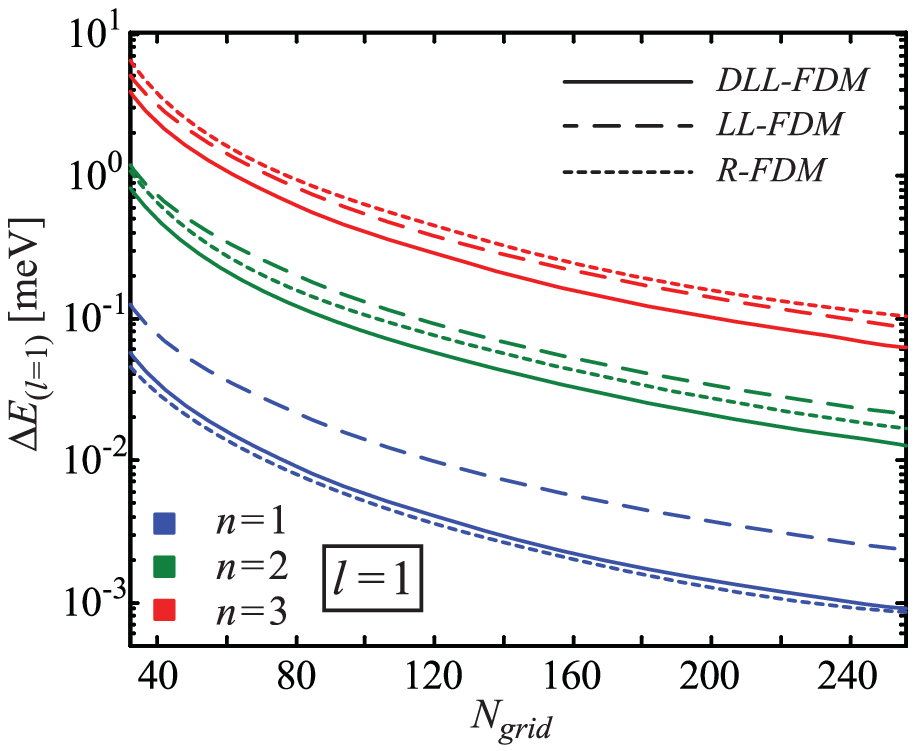}
    \caption{The absolute errors of the $n=1$ (blue lines), $n=2$
    (green lines), and $n=3$ (red lines) states of the orbital
    quantum number $l=1$
    in the rectangular infinite potential well in the freestanding
    quantum wire as function of the number of
    grid points. The {\it DLL-FDM}
    (solid lines), the {\it LL-FDM} (dashed lines)
    and the {\it R-FDM} (dotted lines) schemes are compared.}
    \end{center}
\end{figure}

Finally, we compare in Fig.~5 the errors of the $l=0$ states
computed by the ${\it Se}-{\it FDM}$ and ${\it Sr}-{\it FDM}$
schemes. For all the values of $n$, the absolute errors decay
nearly exponentially with the number of grid points. However, the
${\it Sr}{\it -FDM}$ scheme delivers a few times smaller error
than the ${\it Se}{\it -FDM}$, which could be explained as
follows. The slope at $\rho=0$ is nonzero in the ${\it Se}{\it
-FDM}$ scheme, which increases the eigenenergy with respect to the
value determined by the ${\it Sr-FDM}$ scheme. The error is
smallest for the ground state, as displayed in Fig.~5(a), whereas
the higher states are more oscillatory, and the errors of their
energies, shown in Figs.~5(b) and (c) for $n=2$ and $n=3$, are
much larger than for the $n=1$ state. Nonetheless, in all the
displayed cases in Fig.~5 the errors of the energy levels
continuously decrease with $N_{grid}$. As a matter of fact, the
slopes of the ${\it Se-FDM}$ wave functions at the origin tend to
zero when the grid size increases. It accounts for the convergence
of the ${\it Sr-FDM}$ and ${\it Se-FDM}$ results to each other
when $N_{grid}$ increase, as shown in each panel of Fig.~5.

\begin{figure}
     \begin{center}
       \includegraphics[width=8.6cm]{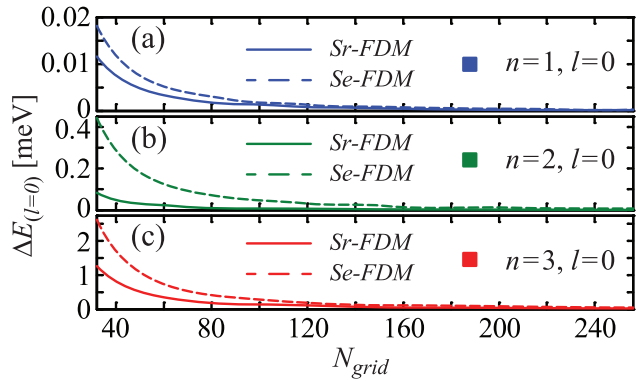}
    \caption{Variation of the absolute errors of the three
    lowest $l=0$ energy levels for the infinite rectangular potential
    in a freestanding quantum wire with the number of grid points for:
    (a) the ground $n=1$ state, (b) the $n=2$ state,
    and (c) the {$n=3$} state. Solid lines are for
    the ${\it S}_r-{\it FDM}$ discretization,
    whereas dashed lines are the results of
    the ${\it S}_e-{\it FDM}$ calculation.}
    \end{center}
\end{figure}

\subsection{The potential of a linear harmonic oscillator in a free-standing quantum wire}
\label{4.2}

The confining potentials of the particles in a quantum wire are
usually assumed to be constant in each material, but vary abruptly
due to band offsets at interfaces between different materials. The
external fields \cite{Sidor2005}, mechanical strain
\cite{Arsoski2013}, interdiffusion \cite{Gunawan2005}, and
self-consistency effects \cite{Proetto1992,Tadic1994} may change
such potential profiles. Self-consistency effects are known to
lead to potentials which may be approximated by parabolas
\cite{Proetto1992,Peeters1990}. Moreover, a linear harmonic
oscillator is an extremely useful model in quantum mechanics.

In the analyzed axially symmetric quantum wire, the model of an
isotropic 2D linear harmonic oscillator is adopted.
\begin{equation}
    V(\rho)=m\omega^2\rho^2/2,
\end{equation}
for which there exist analytical solutions for the eigenenergies,
given by
\begin{equation}
    E_{l,n}=\hbar\omega\left(2n+l+1\right),
\end{equation}
where $n$ denotes the principal quantum number, which is a
nonnegative integer.

\begin{figure}
     \begin{center}
       \includegraphics[width=8.6cm]{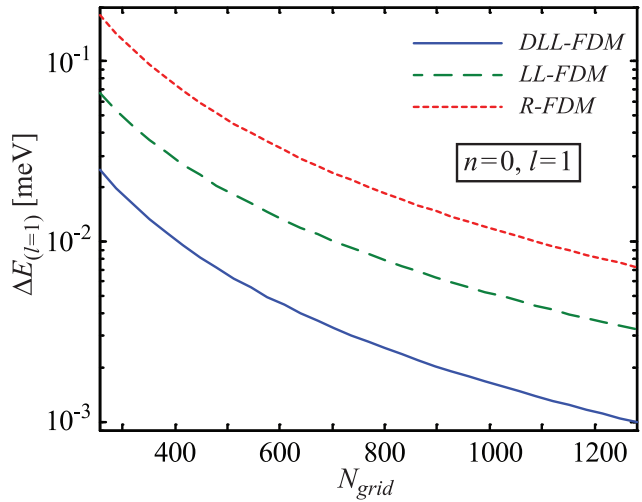}
    \caption{The absolute errors of the $(n=1,l=1)$ energy
    level in the 2D quantum harmonic oscillator computed by
    the {\it DLL-FDM} (solid line), the {\it LL-FDM}
    (long dashed line), and the {\it R-FDM} (short dashed line)
    scheme as function of the number of grid points.}
    \end{center}
\end{figure}

In Fig.~6 we compare the relative errors of the $(l=1,n=0)$ state
for $\hbar\omega=10$ meV, which are made by the {\it DLL-FDM}
(solid lines), the {\it LL-FDM} (dashed lines), and the {\it
R-FDM} (dotted lines) discretization schemes. The numerical
boundary is positioned at $R_{box}=100$ nm, which mimics
experimental conditions in wide quantum wires formed by
lithography and etching \cite{Proetto1992, Youtsey1994} rather than in those
made by the $VLS$ technique. The choice of a wider wire here
largely avoids the problem of reduced accuracy due to the cutoff
of the parabolic potential at $\rho=R_{box}$. However, for the
larger selected domain, to match the range of the step size in
Figs.~2-5, it was necessary to form a larger grid. The case $l=1$
is shown in Fig.~6 because it does not suffer from a low accuracy
close to $\rho=0$ as in the $l=0$ case. Furthermore, the errors of
the higher energy levels are larger than for the lowest energy
state. It is because higher states extend spatially in larger
regions, and therefore are affected more by the inaccuracy of the
wave function close to the numerical boundaries. Hence, we display
in Fig.~6 the energy of only the $n=0$ state.

Fig.~6 shows that out of the three schemes the {\it DLL-FDM}
approach delivers the most accurate energy levels. The energy
levels determined by the {\it R-FDM} substantially deviate from
the results of the {\it DLL-FDM} calculations. As a matter of
fact, the error of the {\it R-FDM} scheme is an order of magnitude
larger than the error of the {\it DLL-FDM}. Moreover, the {\it
R-FDM} gives even less accurate results than the {\it LL-FDM},
which is a consequence of the adaptation in the {\it R-FDM} to the
case when $\psi$ can be approximated by a polynomial $P(\rho)$, as
for the infinite rectangular potential barrier. Hence, the results
of our calculations indicate that the {\it R-FDM} scheme is
inaccurate when used to model the LHO quantum states.

\subsection{The type-Ic confinement potential in a core-shell quantum wires}
\label{4.3}

The third interesting case is a stepwise confinement potential
varying along $\rho$,
\begin{equation}
    V(\rho)=\begin{cases}
                V_c, & \rho\leq R_c \\
                V_{s}, & R_c<\rho< R_s\\
         \infty, & \rho\geq R_s
            \end{cases}.
\label{pot:core:shell}
\end{equation}

It models core-shell quantum wires, where the core ($\rho\leq
R_c$) and the shell ($R_c<\rho< R_s$) are made of different
semiconductors. The effective mass values inside the core and the
shell are different and are equal to $m=m_c$ and $m=m_s$,
respectively, whereas $V_c$ and $V_{s}$ are the conduction band
edges in the core and shell, respectively. Here, we assume that
$V_c<V_{s}$, therefore the electron is confined in the core
(type-Ic confinement). The band offset is defined by
$V_{off}=V_s-V_c$. The analytical solution of the Schr\"odinger
equation in the core is proportional to the Bessel function of the
first kind $J_l(k\rho)$, and inside the shell it is a linear
combination of the modified Bessel functions of the first and the
second kind, $I_l(\kappa\rho)$ and $K_l(\kappa\rho)$,
respectively. Here, $k=\sqrt{2m_cE/\hbar^2}$ and
$\kappa=\sqrt{2m_s(V_{off}-E)/\hbar^2}$. Using the boundary
conditions, $\psi(R_{c}^-)=\psi(R_{c}^+)$,
$(1/m_c)\psi^\prime(R_{c}^-)=(1/m_s)\psi^\prime(R_c^+)$, and
$\psi(R_{s}^-)=0$ results into
\begin{eqnarray}
    \left| \begin{array}{ccc}
        J_l(kR_c) & -I_l(\kappa R_c) & -K_l(\kappa R_c) \\
        \frac{k}{m_c}J'_l(kR_c) & -\frac{\kappa}{m_s}I'_l(\kappa R_c) & -\frac{\kappa}{m_s}K'_l(\kappa R_c) \\
        0 & I_l(\kappa R_s) & K_l(\kappa R_s) \end{array}
        \right|=0,
\end{eqnarray}
wherefrom the eigenenergies are computed. Here, the prime symbol
denotes the first derivative with respect to either $k\rho$ or
$\kappa\rho$.

\begin{figure}
     \begin{center}
       \includegraphics[width=8.6cm]{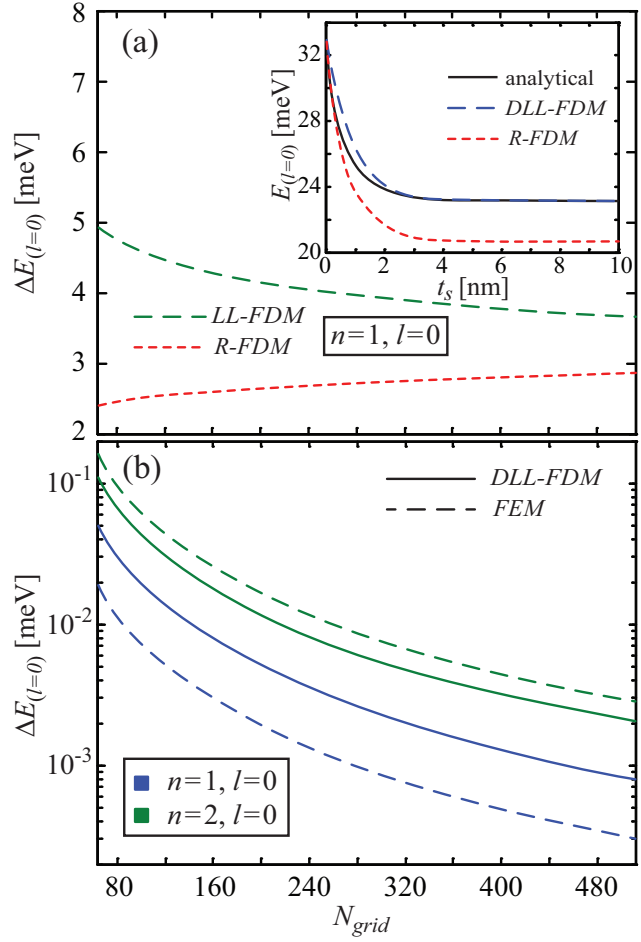}
    \caption{(a) Variation of the absolute errors of the
    ground state $l=0$ eigenenergy for the type-Ic confinement
    potential in the core-shell quantum wire computed by
    the {\it LL-FDM} (green dashed lines) and {\it R-FDM} (dotted
    red lines) schemes with the number of grid points.
    Inset shows variation of the ground state eigenenergy
    with the shell thickness. (b) The absolute errors of
    the {\it DLL-FDM} (solid lines) and the {\it FEM} (dashed lines)
    calculations of the $l=0$ levels: $n=1$ (blue lines) and $n=2$
    (green lines).}
    \end{center}
\end{figure}

To adjust the grid step to the previously analyzed potentials, the
number of grid points is twice as large as in the case of the
infinite well. The analyzed system is the GaAs/(Al,Ga)As
core-shell nanowire, with the core radius $R_c=10$ nm, and the
outer radius of the shell is $R_s=20$ nm. The absolute errors of
the $l=0$ eigenenergies are displayed in Fig.~7, where Fig.~7(a)
compares the {\it LL-FDM} and {\it R-FDM} schemes, and Fig.~7(b)
shows the comparison between the {\it DLL-FDM} scheme and the {\it
FEM}. As for the previous three potentials, the accuracy of the
{\it DLL-FDM} scheme is much better than the accuracy of the {\it
LL-FDM} scheme (compare Figs.~7(a) and (b)). But, opposite to the
LHO, the ground state energy computed by the {\it R-FDM} has a
better accuracy than the {\it LL-FDM}. Yet, the error of the {\it
R-FDM} eigenenergy is much larger than the {\it DLL-FDM} result.
We recall that an estimation of the range where the second
derivative is adapted to the solution of the infinite rectangular
quantum well is needed in the {\it R-FDM} scheme \cite{Rizea2008}.
When the confinement potential in core-shell quantum wires is of
type-Ic, we found that the {\it R-FDM} scheme gives the best
result when the adaptation is done in the whole core, which is the
result displayed in Fig.~7(a).

We also analyzed how the energy levels vary when the shell
thickness $t_s=R_s-R_c$ decreases, which is plotted in Fig.~7(a).
Here, the number of grid points in the core is fixed to 100.
Because the wave function is confined in a larger portion of the
structure, the error of the {\it R-FDM} approximation decreases
when $t_s$ decreases, and therefore its result, shown by the short
dashed curve in the inset of Fig.~7(a) approaches the two other
curves at $t_s=0$. The {\it DLL-FDM} computed energy, displayed by
the long dashed curve, differ negligibly from the analytical
solution when $t_s>2$ nm, whereas for smaller $t_s$ the difference
between the two is larger. A large error of the {\it R-FDM} for
large $t_s$ can be explained by the fact that the solution in the
shell is expressed as a linear combination of the Bessel functions
$K_l$ and $I_l$. Such a combination implies that a higher order
polynomial $P(\rho)$ should be used in the {\it R-FDM} scheme.

Since the results obtained by means of the {\it DLL-FDM} are much
more accurate than both the {\it LL-FDM} and {\it R-FDM}, in
Fig.~7(b) we choose to compare the {\it DLL-FDM} with the {\it
FEM} calculation which employs the linear Lagrange basis and uses
the same uniform grid as the {\it DLL-FDM}. Fig.~7(b) shows the
absolute errors of the two lowest $l=0$ energy levels computed by
the {\it DLL-FDM} (solid lines) and the {\it FEM} (dashed lines)
as function of the number of grid points. The accuracy of the
ground energy level computed by the {\it FEM} is a few times
better than the accuracy of the {\it DLL-FDM}, even though the
lowest order approximations are adopted in both calculations.
Also, the ground energy level obtained by means of the {\it FEM}
exhibits a faster convergence toward the exact value than the {\it
DLL-FDM}. Nonetheless, the accuracy of the {\it DLL-FDM} is
slightly better for the $n=2$ state.

\subsection{The type-Is confinement potential in core-shell quantum wires}
\label{4.4}

Let us now consider the case of the potential as in
Eq.(\ref{pot:core:shell}), but for $V_c>V_{s}$, which assumes that
the electron is mainly confined in the shell (type-Is
confinement). The analytical solution in the core is proportional
to the modified Bessel function of the first $I_l(\kappa\rho)$,
whereas in the shell it is formed as a linear combination of the
Bessel functions of the first and the second kind, $J_l(k\rho)$
and $Y_l(k\rho)$, respectively. Here, $k=\sqrt{2m_sE/\hbar^2}$,
$\kappa=\sqrt{2m_c(V_{off}-E)/\hbar^2}$ and $V_{off}=V_c-V_s$. By
using the same boundary conditions as for the type-Ic confinement,
we derive the equation:
\begin{eqnarray}
    \left| \begin{array}{ccc}
        I_l(\kappa R_c) & -J_l(kR_c) & -Y_l(kR_c) \\
        \frac{\kappa}{m_c}I'_l(\kappa R_c) & -\frac{k}{m_s}J'_l(kR_c) & -\frac{k}{m_s}Y'_l(kR_c) \\
        0 & J_l(kR_s) & Y_l(kR_s) \end{array} \right|=0.
\end{eqnarray}
The energy levels determined from this equation are used as a
reference for assessing the accuracy of the {\it FDM} schemes.

We analyzed the (Al,Ga)As/GaAs core-shell nanowire, whose core and
shell radii are $R_c=3$ nm and $R_s=6$ nm, respectively. The
smaller dimensions are selected here than for the case of the
type-Is confinement because we found almost no difference between
the results of the {\it LL-FDM}, the {\it DLL-FDM}, and the {\it
R-FDM} schemes for $R_c=10$ nm and $R_s=20$ nm. This is a
consequence of the decay of the wave function to a negligible
value inside the core for such a large radius, where the {\it
LL-FDM} was previously demonstrated to lack accuracy.  Thus the
error of both the wave functions and the energy levels computed by
the {\it LL-FDM} approaches the results obtained by the other
methods. For $R_c=3$ nm and $R_s=6$ nm, however, the {\it FDM}
schemes exhibit considerable mutual discrepancy, as Fig.~8 shows.
The most accurate result is again computed by the {\it DLL-FDM}
scheme. The core is here rather thin such that the wave function
does not vanish in the core center, and is proportional to the
modified Bessel function of the first kind $I_l$, which has a
different shape than $J_l$ to which the {\it R-FDM} scheme is
adapted. Therefore, a large discrepancy of the {\it R-FDM} energy
level from the {\it DLL-FDM} result is found. However, the {\it
R-FDM} scheme has a slightly better accuracy than the {\it LL-FDM}
approach. Therefore, the three schemes displayed in Fig.~8 compare
similarly as for the case of type-Ic confinement (compare Figs.~7
and 8).

\begin{figure}
     \begin{center}
       \includegraphics[width=8.6cm]{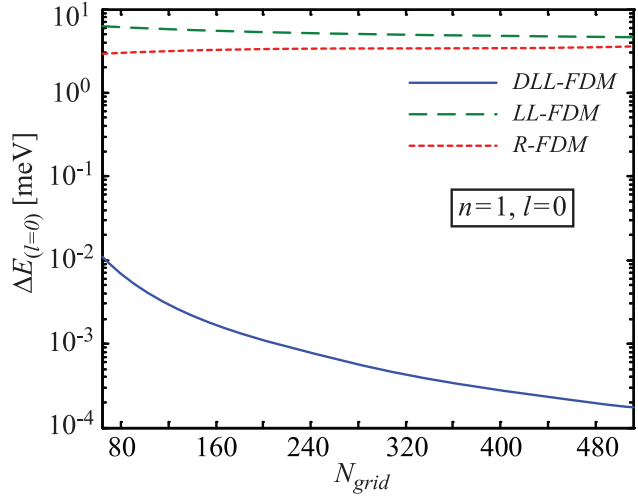}
    \caption{Variation of the absolute errors of the
    $l=0$ ground state eigenenergy for the type-Is
    confinement in the core-shell quantum wire
    with the number of grid points determined by
    the {\it DLL-FDM} (solid blue lines),
    the {\it LL-FDM} (dashed green lines),
    and the {\it R-FDM} (dotted red lines).}
    \end{center}
\end{figure}

\subsection{A note on the PT symmetry of the Hamiltonian}
\label{4.5}

From the analysis presented in previous sections we might recall
that solving the original Schr\"odinger equation brings about a
nonsymmetric Hamiltonian matrix. But, we found that all the
eigenvalues of such a matrix are always real. More specifically,
we found that the solutions of the equations according to the {\it
Sr-FDM} and {\it DLL-FDM} schemes are equal, even though the
Hamiltonian matrices which are constructed by the two schemes have
different symmetry. At this point in our work, we raised the
question about the reasons for the realness of the eigenvalues of
the nonsymmetric Hamiltonian matrix. It needed examining the
solutions of Eq.~(\ref{eq:seqr}) more closely. In order to explain
our findings, we rewrite Eq.~(\ref{eq:seqr}) by changing the
symbol $\rho$ to $x$,
\begin{eqnarray}
    &-&\frac{\hbar^2}{2}\left\lbrack\frac{1}{m^*}\frac{d^2\psi_x(x)}{d x^2}
    +\frac{1}{m^*}\frac{1}{x}\frac{d\psi_x(x)}{dx}+\frac{d}{dx}
    \left(\frac{1}{m^*}\right)\frac{d\psi_x(x)}{dx}
    -\frac{l^2\psi_x(x)}{m^*x^2}\right\rbrack\nonumber\\
    &+&V(x)\psi_x(x)=E\psi_x(x).
    \label{eq:ptx}
\end{eqnarray}
It is the Schr\"odinger equation for $\psi_x$,
\begin{equation}
    H_x\psi_x(x)=E\psi_x(x).
\end{equation}
This equation is easily solved even if $x$ is assumed to be a
Cartesian coordinate. However, the Hamiltonian $H_x$ is manifestly
non-Hermitian,
\begin{equation}
    \int\displaylimits_0^{R_{box}}\psi_{xi}^*(x)H_x\psi_{xj}(x)dx\neq
    \left[\int\displaylimits_0^{R_{box}}\psi_{xj}^*(x)H_x\psi_{xi}(x)dx\right]^*,
\end{equation}
and hence the matrix representation of $H_x$ is non-Hermitian as
well. In the special case of real $\psi_{xi}$ and $\psi_{xj}$ the
last equation points out that the Hamiltonian matrix would be
nonsymmetric. Such a result was previously obtained by either the
{\it Sr-FDM} or the {\it Se-FDM} discretization schemes.

Because the Hamiltonian $H_x$ is non-Hermitian, it may {\it a
priori} turn out that the eigenvalues are non-real. However, $H_x$
satisfies another condition, which guarantees that all its
eigenvalues are real. In order to formulate it, the range of $x$
should be extended to $x<0$. Also, because the wave functions
should satisfy either the Dirichlet or the von Neumann boundary
condition, $V(x)$ and $1/m^*(x)$ should be symmetric in the
extended range of $x$: $V(-x)=V(x)$ and $1/m^*(-x)=1/m^*(x)$. It
is then straightforward to show that $H_x$ commutes with the
product of the space reflection operator ${\cal P}$ and the
time-reversal operator ${\cal T}$ \cite{Bender2007},
\begin{equation}
    [H_x,{\cal PT}]=0.
\label{hx:pt}
\end{equation}
Since $H_x$ satisfies Eq.~(\ref{hx:pt}) it is a ${\cal PT}$
symmetric Hamiltonian \cite{Bender2007}.

If the same boundary conditions are imposed to both
Eqs.~(\ref{eq:derho}) and (\ref{eq:ptx}), they should have the
same spectrum (eigenvalues), despite the fact that $H$ is a
Hermitian operator, whereas $H_x$ is not. Therefore, $H_x$ is a
non-Hermitian operator having real eigenvalues, which is a
property of any ${\cal PT}$ symmetric Hamiltonian. This conclusion
is quite general and does not have any relation to how the
Schr\"odinger equation is solved. It indicates that any method for
solving the Schr\"odinger equation which (mistakenly or
deliberately) treats $\rho$ as a coordinate of the rectilinear
system should deliver real eigenvalues. Hence, if multiplication
by $\rho$ is missing in the matrix elements of $H$, i.e. if they
are computed as $\int\psi_i(\rho)H\psi_j(\rho)d\rho$, the same
eigenvalues are obtained as if the matrix elements are properly
computed as $\int\psi_i(\rho)H\psi_j(\rho)\rho d\rho$. But the
boundary conditions must also be equal for these two calculations
to give the same energy spectrum.

Let us refer back to our comparison of the different {\it FDM} schemes.
The {\it LL} transformed Hamiltonian $\tilde{H}$ in
Eq.~(\ref{eq:trans}) is Hermitian, such that it satisfies,
\begin{equation}
    \int\displaylimits_0^{R_{box}}F_i^*(\rho)\tilde{H}F_j(\rho)
    d\rho=
    \int\displaylimits_0^{R_{box}}F_j^*(\rho)\tilde{H}F_i(\rho)
    d\rho,
\end{equation}
therefore after the {\it LL} transformation, $\rho$ is treated as
a Cartesian coordinate. However, Eq.~(\ref{eq:seqf}) is solved by
applying a different boundary condition than for
Eq.~(\ref{eq:derho}). It makes the results of the {\it LL-FDM}
calculations different from those obtained by means of the {\it
S-FDM} schemes. On the other hand, the {\it DLL-FDM} scheme
produces a symmetric Hamiltonian matrix and gives the same energy
levels as the {\it Sr-FDM} scheme. It is because equivalent
boundary conditions are implemented in the two methods. Also, the
Hamiltonian obtained by the adaptive calculation of Rizea et al.
\cite{Rizea2008} is ${\cal PT}$ symmetric. It explains why this
procedure delivers all real eigenvalues even though the
constructed Hamiltonian matrices are nonsymmetric.

The conclusions about the energy spectrum that have just been
derived from the ${\cal PT}$ symmetry property of the Hamiltonian
$H$ are quite general and are therefore not related to a specific
numerical method for solving the effective-mass Schr\"odinger
equation. Hence, they are valid for computations by the
finite-element method and the method of expansion, for example.
However, among all the methods, those which produce symmetric
Hamiltonian matrices offer much more efficient numerical solutions
of the eigenproblem due to the important advantages of symmetric
matrices: abundance of software for their numerical
diagonalization, generally faster diagonalization procedures, and
reduced requirements for memory storage. Nonetheless, for a
successful application of any such a method it is important to
properly implement the boundary conditions, as is done by the {\it
DLL-FDM} scheme.

\section{Conclusions}
\label{5}

An efficient discretization scheme to solve the Schr\"{o}dinger
equation in cylindrical coordinates is devised, and applied to
compute the energy levels for a few potentials in axially
symmetric quantum wires and quantum dots. It is constructed by the
{\it FDM} discretization of the original Schr\"odinger equation,
and the subsequent transformation applied to the difference
equation. Thus the Hamiltonian matrix is symmetrized in discrete
space. The scheme is demonstrated to lead to an improved accuracy
of the numerical solution close to the inner boundary of the
numerical domain. Also, the Hamiltonian matrices which are
constructed by the scheme are symmetric, and are therefore more
efficiently diagonalized than the nonsymmetric matrices formed by
a direct application of central differences to the Schr\"odinger
equation. We infer that the proposed discretization scheme could
be applied to other differential equations in cylindrical
coordinates which contain the $\rho$-dependent part of the
Laplacian. In addition to the symmetry of the Hamiltonian the
scheme is shown to have an improved accuracy with respect to other
methods close to $\rho=0$. Moreover, it is shown to compare
favorably well with the results of the finite-element
calculations. The coincidence of the results obtained by some {\it
FDM} schemes is explained by the ${\cal PT}$ symmetry of the
Hamiltonian, and the large error of some of them is found to be
mainly a consequence of the peculiar boundary conditions.

\section{Acknowledgements}

This work was supported by the Ministry of Education, Science,
and Technological Development of Serbia (project III 45003) and
the Fonds Wetenschappelijk Onderzoek (Belgium).

\bibliographystyle{elsarticle-num}



\end{document}